\documentclass[iop,twocolumn,numberedappendix]{aastex62}
\newcommand{\be}{\begin{equation}}
\newcommand{\ee}{\end{equation}}

\usepackage{multirow}
\usepackage{natbib, xfrac}
\usepackage{enumerate}
\usepackage{graphicx}
\usepackage{amsmath, amssymb}
\usepackage{courier}
\usepackage{color}
\usepackage{hyperref}
\usepackage{float}

\begin{document}
\title{Constraining Gas-Phase Carbon, Oxygen, and Nitrogen in the IM Lup Protoplanetary Disk}
\shorttitle{C, N, and O Abundances in the IM Lup disk}
    \shortauthors{Cleeves et al.}
    
\author{L. Ilsedore Cleeves}
\altaffiliation{Hubble Fellow}
\affiliation{Harvard-Smithsonian Center for Astrophysics, Cambridge, MA 02138}
\affiliation{Astronomy Department, University of Virginia, Charlottesville, VA 22904, USA}

\author{Karin I. {\"O}berg}
\affiliation{Harvard-Smithsonian Center for Astrophysics, Cambridge, MA 02138}

\author{David J. Wilner}
\affiliation{Harvard-Smithsonian Center for Astrophysics, Cambridge, MA 02138}

\author{Jane Huang}
\affiliation{Harvard-Smithsonian Center for Astrophysics, Cambridge, MA 02138}

\author{Ryan A. Loomis}
\altaffiliation{Jansky Fellow of the National Radio Astronomy Observatory} 
\affiliation{Harvard-Smithsonian Center for Astrophysics, Cambridge, MA 02138}
\affiliation{National Radio Astronomy Observatory, Charlottesville, VA 22903, USA}

\author{Sean M. Andrews}
\affiliation{Harvard-Smithsonian Center for Astrophysics, Cambridge, MA 02138}

\author{V. V. Guzman}
\affiliation{Joint ALMA Observatory (JAO), Alonso de C\'ordova 3107 Vitacura, Santiago de Chile, Chile}

\begin{abstract}
We present new constraints on gas-phase C, N, and O abundances in the molecular layer of the IM Lup protoplanetary disk. Building on previous physical and chemical modeling of this disk, we use new ALMA observations of C$_2$H to constrain the C/O ratio in the molecular layer to be $\sim0.8$, i.e., higher than the solar value of $\sim0.54$. We use archival ALMA observations of HCN and H$^{13}$CN to show that no depletion of N is required (assuming an interstellar abundance of $7.5\times10^{-5}$ per H).  These results suggest that an appreciable fraction of O is sequestered in water ice in large grains settled to the disk mid-plane. Similarly, a fraction of the available C is locked up in less volatile molecules. By contrast, N remains largely unprocessed, likely as N$_2$. This pattern of depletion suggests the presence of true abundance variations in this disk, and not a simple overall depletion of gas mass.  If these results hold more generally, then combined CO, C$_2$H, and HCN observations of disks may provide a promising path for constraining gas-phase C/O and N/O during planet-formation. Together, these tracers offer the opportunity to link the volatile compositions of disks to the atmospheres of planets formed from them.
 \end{abstract}


\section{Introduction} 
Gas-rich circumstellar disks around young stars provide a window to study the materials that are incorporated into forming planetary systems. Chemistry and dynamics in the disk can shift the balance of gas- versus ice-phase volatiles containing, e.g., carbon, nitrogen, and oxygen. In the core accretion paradigm \citep{pollack1996}, those materials that end up as rocks or ices are incorporated into the ``solid'' planetesimals, while the remaining gas can be accreted into natal planets' atmospheres. Correspondingly, it is essential to understand the form volatiles containing carbon, nitrogen, and oxygen take in the disk spatially and over time, to understand the initial elemental compositions of forming planets \citep[e.g.,][]{oberg2011co,piso2016,cridland2016,cridland2017}. 

Many processes can alter the gas versus solid abundances in the disk, including snow lines \citep[e.g.,][]{oberg2011co}, chemistry \citep{bergin2014far,furuya2014,eistrup2016,schwarz2018}, mixing and/or diffusion of gas \citep[e.g.,][]{semenov2011,kama2016}, and redistribution of ices as dust grows and evolves \citep[e.g.,][]{hogerheijde2011,piso2016,krijt2016,oberg2016}.  To date, observations of various carbon and oxygen carriers have suggested a substantial ``missing'' volatile mass within the disk molecular layer as traced by CO in the submillimeter \citep{favre2013,cleeves2015tw,zhang2017} and H$_2$O vapor in the far-infrared with {\em Herschel} \citep{bergin2010,hogerheijde2011,du2017}, though see also \citet{kamp2013} regarding model dependencies. For the few disks where estimates for both exist, more water in the disk surface is ``missing'' than CO when compared to interstellar abundances, and both have abundances lower than what simple desorption (thermal and non-thermal) models predict \citep[e.g., in TW Hya's disk;][]{hogerheijde2011,favre2013,schwarz2016,kama2016,cleeves2015tw}. 

Absolute elemental abundances are often difficult to estimate due to uncertain hydrogen disk masses. In this context, relative elemental abundances, such as C/O and N/O are promising avenues toward robustly characterizing disk volatile compositions.
Recently, \citet{bergin2016} reported bright hydrocarbon rings of C$_2$H and $c-$C$_3$H$_2$ in the TW Hya \citep[see also][]{kastner2015} and DM Tau disks. Using chemical models, \citet{bergin2016} found the abundances of these small hydrocarbons were especially sensitive to the gas-phase C/O ratio of the disk. The observations required high C/O values, $>1$,  to reproduce the observed line intensities \citep[see also][]{du2015}. 
Such prospects for measuring C/O in disks are now especially exciting as we enter an era where the elemental compositions, including C/O, of exoplanets' atmospheres \citep[e.g.,][]{madhusudhan2011,kreidberg2014,macintosh2015,bonnefoy2016,lavie2017}.

In contrast to carbon and oxygen abundance estimates, there are few constraints on total nitrogen abundances or N/O ratios in disks owing in large part to the difficulty of observing  the likely primary nitrogen carrier,  N$_2$, \citep[e.g.,][and references therein]{schwarz2016}. Abundant nitrogen bearing species, such as N$_2$H$^+$ and HCN, are sensitive to other disk parameters than total N abundance, such as temperature structure, carbon abundance, and ionization rate. As such, interpreting these species in the context of bulk nitrogen abundance requires detailed knowledge of the source.

In this work, we constrain the carbon, nitrogen, and oxygen content of the warm molecular layer in the IM Lup protoplanetary disk using results from our previous study of CO and its isotopologues \citep{cleeves2016im} and new and archival ALMA observations of C$_2$H and HCN and H$^{13}$CN. The solar mass star IM Lup harbors a massive gas rich disk, $M_{\rm disk} \sim0.1-0.2$~M$_{\odot}$  based upon continuum (SED and resolved millimeter images) and CO multi-isotopologue multi-line data presented in \citep{cleeves2016im}. The source is relatively young at an age of $0.5-1$~Myr \citep{mawet2012}.

In \citet{cleeves2016im}, we found that IM Lup's CO is under-abundant by a factor of $\sim20$ compared to an interstellar CO abundance of $1.4\times10^{-4}$ per H based upon the dust-inferred disk mass. For comparison, this younger object appears to be ``missing'' less CO than the older TW Hya system, whose CO abundance is $\sim3-5\times$ less abundant than IM Lup \citep[e.g.,][]{favre2013}.   However, based on this data alone it is difficult to tell whether the observed ``missing'' gas-phase CO is a result of missing carbon or missing oxygen or both; or, alternatively, missing total gas mass compared to dust. In the present paper, we explore what C/O and N/O abundance ratios are required in the disk's warm molecular layer to reproduce the observed C$_2$H, HCN, and HC$^{13}$N line intensities.

\section{Observations}\label{sec:obs}

\begin{figure*}[]
\begin{centering}
\includegraphics[width=1.0\textwidth]{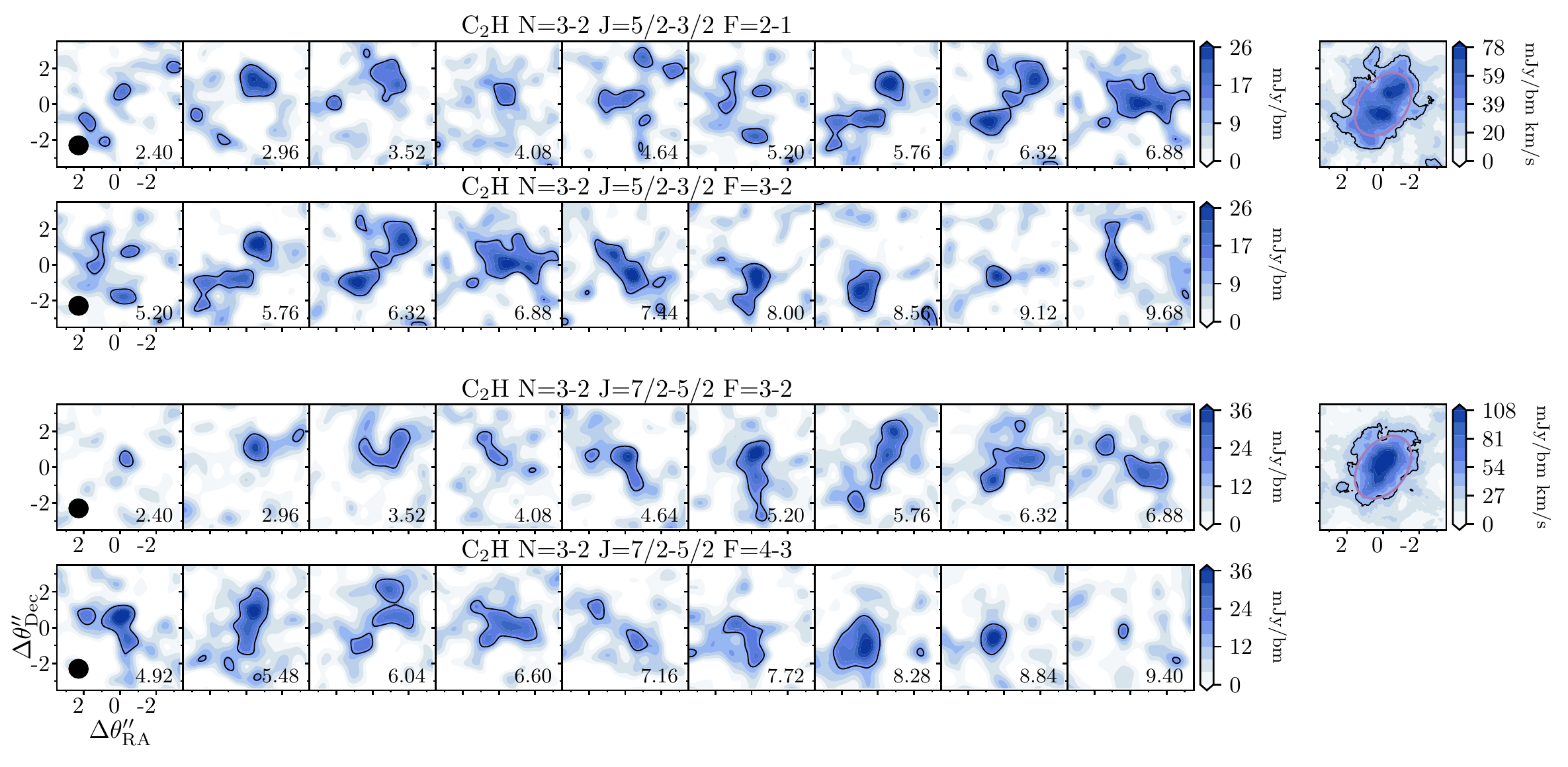}
\caption{Channel maps of C$_2$H $N=3-2$ for the a)  $J=\sfrac{5}{2} - \sfrac{3}{2}$ pair and b) $J=\sfrac{7}{2} - \sfrac{5}{2}$ pair, channel averaged by a factor of two for clarity. THe solid contour line indicates 3$\sigma$. The beam is in the lower left corner. The VLSR in km~s$^{-1}$ is indicated in the bottom right corner of each panel.  The purple ellipse on the right panels indicates the scale of the millimeter thermal dust emission, $R=313$ AU \citep{cleeves2016im}. \label{fig:allchannel}}
\end{centering}
\end{figure*} 

\subsection{ALMA Observations and Data Reduction}\label{sec:data}

\begin{figure}[]
\begin{centering}
\includegraphics[width=0.34\textwidth]{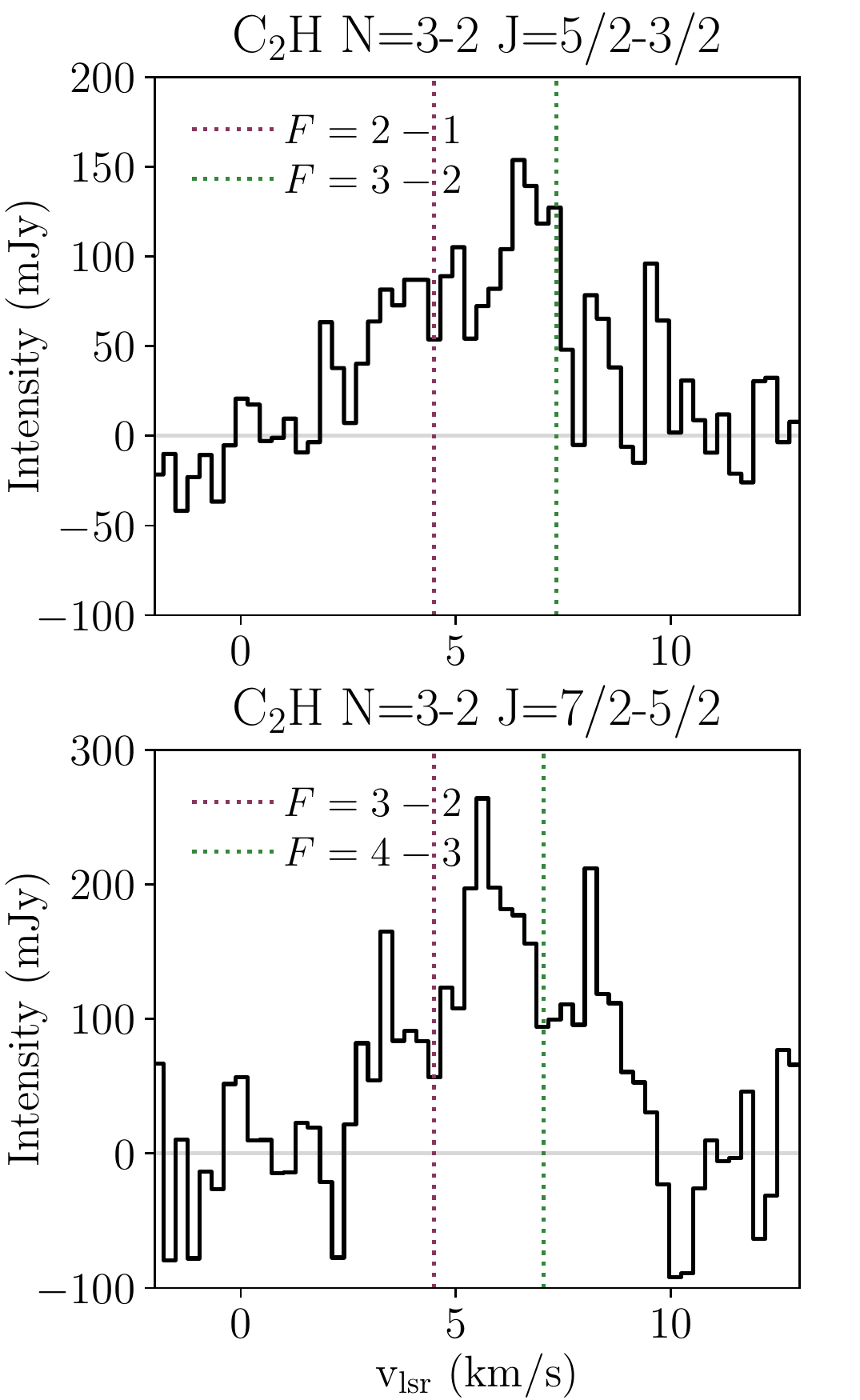}
\caption{C$_2$H $N=3-2$ spectra extracted from a circular 4'' mask. The dotted line indicates the line centers for each pair in the rest frame of the $F=2-1$ (top) and $F=3-2$ (bottom) transitions. \label{fig:spec}}
\end{centering}
\end{figure} 

IM Lup was observed as part of a survey of nitrogen isotope chemistry and deuterium chemistry presented in \citet{guzman2017} and \citet{huang2017}, respectively. Despite this source being one of the most gas-rich disks and bright in HCN $J=3-2$ \citep[see also][]{oberg2011discsii}, the H$^{13}$CN $J=3-2$ line was not detected at a per channel RMS of 3.6 mJy per beam in 0.5 km s$^{-1}$ channels \citep{huang2017}. The observations of 
HCN $J=3-2$ and H$^{13}$CN $J=3-2$ are presented in \citet{huang2017} \citep[see also][]{guzman2017} and are not reproduced here.

The C$_2$H $N=3-2$ hyperfine complex was observed with ALMA as part of the Cycle 3 2015.1.00964.S program (PI: {\"O}berg) on 01 May 2016 with 41 antennae for 12~minutes on source.
The observations were calibrated by ALMA/NAASC staff using J1517-2422 for the bandpass, J1610-3958 for the phase and amplitude, and Titan for the flux calibration. We performed one additional round of phase self-calibration using the continuum within the spectral window containing the line. For the self-calibration, we use a solution interval of 30 seconds and average both polarizations.  The continuum is estimated from line-free channels within the same spectral window, which is then subtracted in the $uv$-plane to obtain continuum-subtracted images. We detect two blended hyperfine pairs of C$_2$H $N=3-2$, the $J=\sfrac{5}{2} - \sfrac{3}{2}$ $F=2-1$ and $F=3-2$ pair and the $J=\sfrac{7}{2} - \sfrac{5}{2}$ $F=3-2$ and $F=4-3$ pair, see Table~\ref{tab:obs}. While the pairs are blended in velocity space, the inclined orientation of the Keplerian disk causes the emission from each of the pairs to be mostly spatially resolved on the sky. Images are generated using CASA 4.7 \citep{mcmullin2007} and the clean task.  Figure~\ref{fig:allchannel} shows velocity-averaged channel maps and moment-0 maps tapered to 1'' to improve signal to noise. The moment-0 maps show the integrated combined emission from each pair after clipping the individual channels below 1$\sigma$. The disk-integrated spectrum of  the observed C$_2$H transitions is shown in Figure~\ref{fig:spec} for each of the hyperfine pairs integrated within a 4'' radius circular aperture. Table~\ref{tab:obs} provides the C$_2$H $N=3-2$ disk integrated fluxes within this aperture, with errors estimated from a 4'' circular aperture in line-free portions of the spectrum combined with 10\% calibration uncertainty added in quadrature.

\begin{deluxetable}{ccc}
\tablecaption{C$_2$H $N=3-2$ Line Observations \label{tab:obs}}
\tabletypesize{\footnotesize}
\tablehead{{Transition } & Rest                                          &  Disk-Integrated   \\
                                     &Freq.&                                                   Flux Density (Blended)                \\
                                                   &     [GHz]      &                           [Jy km s$^{-1}$]         } 
\startdata
 $J=\sfrac{5}{2} - \sfrac{3}{2}$ $F=2-1$ & 262.067 &  $0.56  \pm 0.08$                  \\
$J=\sfrac{5}{2} - \sfrac{3}{2}$ $F=3-2 $& 262.065 &         \\
$J=\sfrac{7}{2} - \sfrac{5}{2}$ $F=3-2$ &  262.006 &   $0.76\pm 0.11$               \\
$J=\sfrac{7}{2} - \sfrac{5}{2}$ $F=4-3$ &  262.004   &                
\enddata
\tablecomments{Uncertainties are quadrature combined RMS scatter and 10\% calibration uncertainty.} 
\end{deluxetable}

\section{Methods}\label{sec:methods}
The following sections describe our methods to constrain carbon, oxygen, and nitrogen abundances in IM Lup's disk using chemical models to find what elemental abundances best reproduce the  C$_2$H, HCN, and H$^{13}$CN observations, while also remaining consistent with the previous CO constraints \citep{cleeves2016im}. 

\subsection{Disk Physical Model}\label{sec:physmod}
We use the underlying disk structure derived in \citet{cleeves2016im}, which we summarize here. The disk surface density follows the self-similarity solutions of \citet{lyndenbell1974}, with an inner power law and outer exponential taper in the disk surface density. The disk is vertically extended and flared, with a gas scale height of 12 AU at 100 AU. The millimeter grains form a thinner layer (25\% of the gas scale height) and contain 99\% of the total dust mass, $M_{\rm dust} = 0.0017$ M$_\odot$. We furthermore assume small grains follow the gas distribution and allow the millimeter grains to have a separate power law distribution with a cut off at the millimeter emission edge at $313$ AU. The small grains provide the greatest surface area per unit mass, and also fill most of the disk volume, and thus are the most important for the grain surface chemistry in the observable layers of the disk ($\gtrsim1$ scale height). In the model, we track the total surface area per unit volume throughout the disk as an input to the chemical calculations. 

The gas and dust temperature structures are fixed, where the dust temperature is calculated assuming radiative equilibrium with a stellar $T_{\rm eff}=3900$~K and $R_*=2.5~R_\odot$ \citep{pinte2008,cleeves2016im}. We note \citet{alcala2017} provides updated stellar parameters for IM Lup; however, to within the uncertainties the new values are similar, and so we have chosen to keep these values fixed to more readily compare to previous work. 
The gas temperature deviates from the dust temperature in the disk upper layers where there is both FUV heating from the central star and the external radiation field, where we fix the external radiation field to be $G_0=4$ Habing \citep{haworth2017,pinte2018,cleeves2016im}.

For the stellar high energy radiation field, we use the UV spectral template of TW Hya normalized to the observed {\it Swift} UVM2 flux density from IM Lup, and the ``quiescent'' X-ray template presented in \citet{cleeves2013a} normalized to the observed integrated X-ray luminosity provided in \citet{gunther2010} of $4.3\times10^{30}$~erg~s$^{-1}$. The wavelength dependent radiation transport is calculated for both UV and X-rays using the code of \citet{bethell2011u}. 

\subsection{Chemical Modeling Procedure}\label{sec:chemmod}
We calculate the 2D chemical abundances of C$_2$H and HCN using a time-evolving gas-grain model first presented in \citet{fogel2011} and updated and expanded in \citet{cleeves2014par}. The chemical model takes into account the spatial changes in dust surface area per unit volume as described in Section~\ref{sec:modupd} and Appendix~\ref{app:grains}. The simulations are run over 0.5~Myr, corresponding to the lower age limit of IM Lup. However, we reach steady state in the warm molecular layer (the region probed by the observations) well before this time and the results are not affected by this choice. 
Therefore, we find this assumption does not significantly impact our results. We use a non-deuterated chemical network with 5974 reactions and 638 species. We do not take into account carbon isotope chemistry. For the H$^{13}$CN $J=3-2$ line radiation transfer, we simply take the model HCN abundance and assume an isotope ratio of $\rm ^{12}C/^{13}C=70$ \citep{prantzos1996}.

The two main variables considered in the modeling are:
\begin{enumerate}
\item the initial C, N, and O abundances (Section~\ref{sec:modabun})
\item  the cosmic ray ionization rate, $\zeta_{\rm CR}$.
\end{enumerate}
Based on  theory \citep{cleeves2013a} and observations of the TW Hya disk \citep{cleeves2015tw}, the typical cosmic ray ionization in disks may be low, $\zeta_{\rm CR}\le2\times10^{-19}$ s$^{-1}$ per H$_2$. These low rates may be related to magnetized wind modulation or deflection by tangled magnetic fields within the disk  \citep{cleeves2013a}. We consider models that have a cosmic ray ionization rate similar to TW Hya and values approximately one order of magnitude higher and one order of magnitude lower. These models correspond to the ``T Tauri minimum modulation'' (ttm; $\zeta_{\rm CR}=1.0\times10^{-20}$ s$^{-1}$ per H$_2$), ``Solar System maximum'' (ssx; $\zeta_{\rm CR}=2.0\times10^{-19}$ s$^{-1}$ per H$_2$), and the ``Solar System minimum'' (ssm; $\zeta_{\rm CR}=1.3\times10^{-18}$ s$^{-1}$ per H$_2$) from \citet{cleeves2013a}.

\subsubsection{Model updates}\label{sec:modupd}
For the present study, the chemical code has been updated in three ways. First, we have improved the grain surface chemistry to provide more ``realistic'' reaction rates. In the past, we have allowed all ice at a given time to participate in grain-surface chemistry, since our code does not treat the multi-layered nature of the ice like the models of, for example, \citet{hasegawa1993}, \citet{vasyunin2013}, \citet{garrod2013}, and \citet{furuya2016}. As such, without taking layered ices into account, we had implicitly enhanced the efficiency of grain surface chemistry, since more realistically we expect primarily the ice surface to be reactive. To approximate this behavior, we multiply the reaction rates by (number of monolayers)$^{-1}$,
except for atomic H, which we expect to exist mainly as part of the reactive surface. Incorporating this effect is an especially important addition given that most disks appear to have a large amount of settled dust mass, i.e., have a deficit of small grains in their surface, including IM Lup. Essentially, at a fixed volatile abundance, the ice-coating on a given grain can become ``thicker'' with decreasing total grain-surface area per volume, making less of the ice mobile and reactive. 

The second chemical code update is to include a temperature-dependent sticking coefficient as described in \citet{he2016}, rather than assuming perfect sticking for all species. We adopt the generic fit from \citet{he2016} Table 1, except for gas-phase water, where we still assume perfect sticking due to its highly polar nature. 

The third model update is the incorporation of N$_2$ self-shielding using the \citet{li2013} shielding functions and the vertically calculated H$_2$ and N$_2$ column densities. Without the latter, we  over-predict the atomic N density and under-predict the N$_2$ density.

\subsubsection{Model abundances}\label{sec:modabun}
\begin{deluxetable}{llll}[bh!]
\tablecolumns{4}
\tablewidth{0pt}
\tablecaption{Fixed chemical abundances besides CO/C$^+$/H$_2$O ice relative to total H atoms. \label{tab:other}}
\tabletypesize{\footnotesize}
\tablehead{{Molecule} & Abundance & {Molecule} & Abundance }
\startdata
H$_2$ & $5.00\times10^{-1}$ & He & $1.40\times10^{-1}$ \\
N$_2$ & $3.75\times10^{-5}$  & CS & $4.00\times10^{-9}$ \\
SO & $5.00\times10^{-9}$ & HCO$^+$ & $9.00\times10^{-9}$ \\
H$_3^+$ & $1.00\times10^{-8}$ & C$_2$H & $8.00\times10^{-9}$ \\
Si$^+$ & $1.00\times10^{-9}$ & Mg$^+$ & $1.00\times10^{-9}$ \\
Fe$^+$ & $1.00\times10^{-9}$ &  
\enddata
\tablecomments{Note: N$_2$ abundance varied in Section~\ref{sec:hcnfit}.} 
\end{deluxetable}

The primary goal of this work is to constrain the total amount of volatile carbon, nitrogen, and oxygen in the upper layers of the disk, which  may not have solar or interstellar bulk composition. Furthermore, we wish to remain agnostic to the specific process leading to deviations in the bulk abundances from interstellar values. To accomplish this, we take the simple approach of adjusting the initial chemical abundances in our models to explore a range of possible total C/H, O/H, and N/H abundances to jointly reproduce the ALMA observations of C$_2$H and HCN within our \citet{cleeves2016im} model framework.

We have examined the models to ensure that they reach a pseudo-steady state for the species of interest in the warm molecular layer between radii of 20 and 300 AU within a relatively short timescale ($\sim10^3$ years, or 0.2\% of the simulation time). Essentially, the chemistry in this layer quickly re-adjusts to the local conditions and is most sensitive to the bulk C, N, and O content rather than the details of the initial abundance profile. This feature of the warm molecular layer chemistry allows us to constrain the abundances without needing the  full (and uncertain) chemical and physical history of the gas. To confirm this behavior, we have tested models where we move the carbon, oxygen, and nitrogen into different carriers, e.g., N$_2$ versus N versus NH$_3$ and achieve similar output abundances in the warm molecular layer ($z/r\gtrsim0.2$) to within a few percent.

The abundances of the species that are  not varied are listed in Table~\ref{tab:other}, and are motivated by molecular cloud model abundances \citep{fogel2011,cleeves2016im}. Species whose abundances are altered in the initial conditions are in Table~\ref{tab:ini}. The baseline water abundance has been updated to reflect the high-end of interstellar water ice measurements, $\chi({\rm H_2O_{ice}}) = 8\times10^{-5}$ per H \citep{boogert2015}, and thus relative depletion factors  in water reported here may be higher if the intrinsic interstellar water content is higher, e.g., hidden in larger interstellar grains \citep[i.e.][]{vandishoeck2014}.

The starting point of our depletion models is our 2016 paper which revealed CO to be under-abundant by a factor of 19 relative to an interstellar CO abundance of $1.3\times10^{-4}$ per H \citep{ripple2013}, updated from previously $1.4\times10^{-4}$ per H. For each abundance mixture, we require there to be sufficient elemental C and O to be able to produce a CO abundance of $7\times10^{-6}$, but not excess. For example, the carbon abundance can be $7\times10^{-6}$ per H, and the oxygen abundance can be this value or greater (carbon-limited models). Similarly, oxygen can have the same $7\times10^{-6}$ per H abundance, but with equal or excess carbon (oxygen-limited models). For the nitrogen depletion factors, our primary carrier is N$_2$, and to  simulate nitrogen ``removal'' we directly reduce the initial N$_2$ abundance. 

\begin{deluxetable}{cccc}[b!]
\tablecolumns{4}
\tablewidth{0pt}
\tablecaption{Abundances of key C and O species. \label{tab:ini}}
\tabletypesize{\footnotesize}
\tablehead{ C/O & C$^+$ & CO & H$_2$O(gr)}
\startdata
 0.08&   0.0 & $7\times10^{-6}$ &  $8\times10^{-5}$ \\  
  0.47&  0.0 & $7\times10^{-6}$ &  $8\times10^{-6}$ \\
   0.64&    0.0 & $7\times10^{-6}$ &  $4\times10^{-6}$ \\ 
   0.81&    0.0 & $7\times10^{-6}$ &  $1.6\times10^{-6}$ \\ 
   0.9&    0.0 & $7\times10^{-6}$ &  $8\times10^{-7}$ \\ 
   0.95&    0.0 & $7\times10^{-6}$ &  $4\times10^{-7}$ \\ 
   1.0&    $3.5\times10^{-6}$ & $3.5\times10^{-6}$ &  $3.5\times10^{-6}$ \\ 
   1.86&    $9.5\times10^{-6}$ & $3.5\times10^{-6}$ &  $3.5\times10^{-6}$ \\    
    3.71&    $2.25\times10^{-5}$ & $3.5\times10^{-6}$ &  $3.5\times10^{-6}$ 
\enddata
\tablecomments{Models with C/O $\le 1$ are carbon limited, while models with C/O $\ge1$ are oxygen limited. In the latter case, the oxygen is split between CO and H$_2$O(gr).} 

\end{deluxetable}

Even though the chemical reprocessing timescales are short in the upper layers ($z/r\ge0.25$) of the disk, we have nonetheless attempted to create ``realistic'' initial compositions rather than purely atomic. As seen in Table~\ref{tab:ini}, the oxygen and carbon are primarily in H$_2$O, CO and C$^+$ when needed. This choice does not affect the main goals of this study (reproducing the observables), but results in more realistic midplane chemical abundances.
%%%%%%%%%%%%%%%%%%%%%%%%%%%%%%%%%%%%%%%%%%%%%%%%%%%%%%%

\subsection{Model - Observation Comparison}\label{sec:linemod}
We simulate the C$_2$H $N=3-2$, HCN $J=3-2$, and H$^{13}$CN $J=3-2$ observations using the non-LTE radiative transfer code LIME v1.8\footnote{\url{https://github.com/lime-rt/lime}} \citep{brinch2010}. We use the collisional rates input files provided by the Leiden LAMDA database \citep{schoier2005}. All calculations take into account full non-LTE radiation transfer. The HCN and H$^{13}$CN collision rate data is assumed to be the same and sourced from \citet{green1974} and does not include hyperfine structure. The C$_2$H collision rate data is from \citet{spielfiedel2012}. The non-LTE analysis is especially important given that the C$_2$H is abundant in our models across a wide range of densities, down to $n_{\rm H}\sim10^5$ cm$^{-3}$. We simulate each of the C$_2$H $J=\sfrac{5}{2} - \sfrac{3}{2}$ and $J=\sfrac{7}{2} - \sfrac{5}{2}$ line pairs together since their emission covers the same frequency space, even if the emission is not blended spatially. While the CO gas extends out as far as $\sim1000$~AU, the HCN and C$_2$H and HCN extend to about $\sim500$ AU, and so we limit our emission models to this radius to only constrain the abundances where C$_2$H is well-detected, and discuss the implications of this in Section~\ref{sec:cono}. 

The line and millimeter continuum data are modeled in LIME simultaneously as the millimeter continuum opacity is known to be high in this source, especially in the inner disk \citep{cleeves2016im}. The continuum at these wavelengths is is entirely midplane dominated, in a thin vertical layer. We have not attempted to adjust the inner disk opacity as in \citet{cleeves2016im}, and note that the main effect would be to reduce the observable line flux from the inner $R\lesssim40$~AU, where the C$_2$H indeed shows an inner depression (see Figure~\ref{fig:allchannel}).

The input gas velocities include Keplerian rotation around a solar mass star, thermal broadening, and a fixed turbulent velocity of 100~m~s$^{-1}$. The final spectral resolution of the C$_2$H, HCN, and H$^{13}$CN simulations is 0.28, 0.28, and 0.35 km~s$^{-1}$, but we simulate the cubes at $40\times$ higher spectral resolution than observed and average down to take into account blurring due to channel averaging. The simulations assume a fixed distance of 161~pc \citep[][DR1]{gaia}, and a fixed inclination of the disk midplane of $48^\circ$ \citep{cleeves2016im}. The updated DR2 distance is $158\pm3$ pc, which is sufficiently consistent with the DR1 value, and so we have kept the DR1 distance for ease of comparison.

The models are compared to the ALMA data in the visibility plane, where the \texttt{vis\_sample} package \citep{loomis2018}\footnote{\url{https://github.com/AstroChem/vis_sample}} is used to sample the LIME channel maps at the same spatial frequencies as were observed to create the model visibilities. To assess the goodness-of-fit for models, we compare the $\chi^2$ between the observed and simulated continuum subtracted visibilities. The total model grid size is nine values of C/O and three values of $\zeta_{\rm CR}$ for 27 models total.

\section{Results}\label{sec:results}
\subsection{Chemical Model Results}\label{chemresults}
\begin{figure*}
\begin{centering}
\includegraphics[width=1\textwidth]{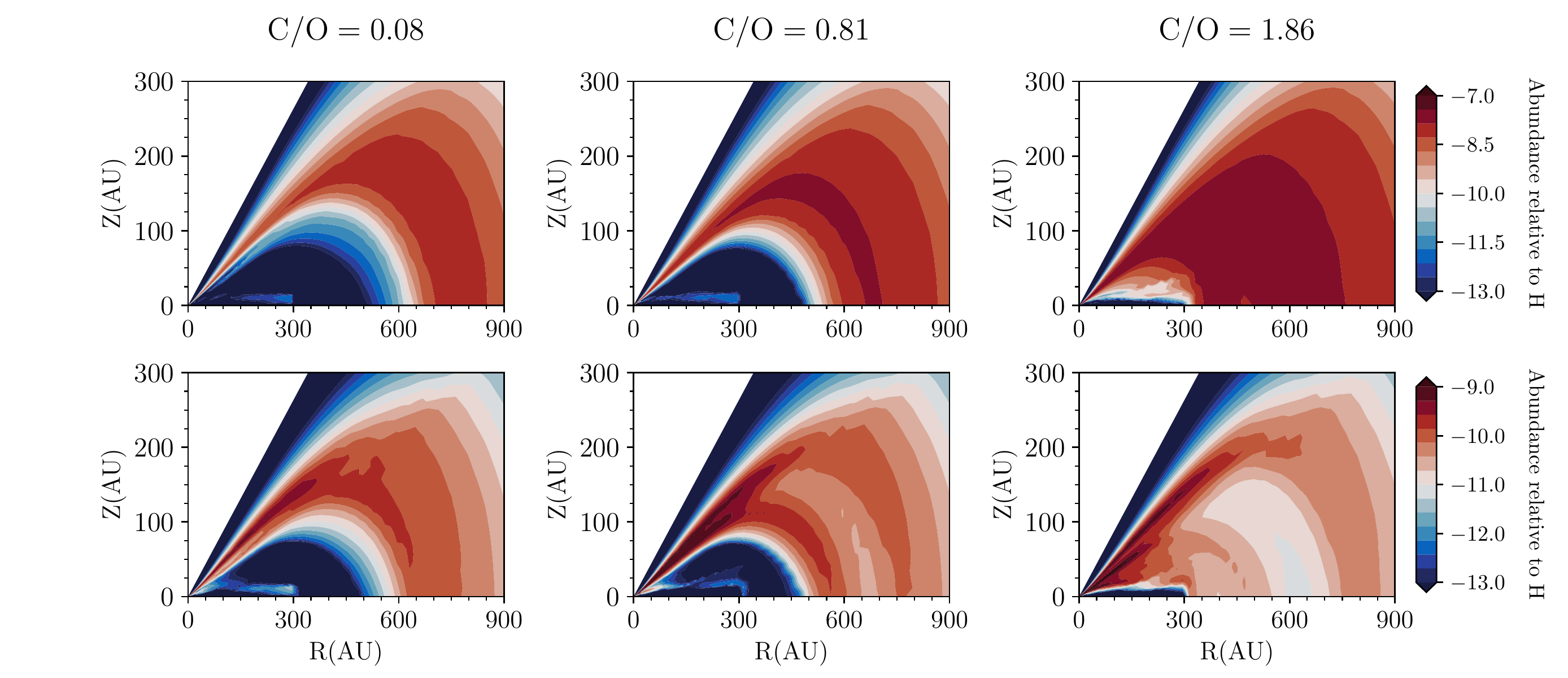}
\caption{C$_2$H (top) and HCN (bottom) 2D abundances for different C/O ratios as labeled at the top of each column at an intermediate cosmic ray ionization rate of $\sim10^{-19}$~s$^{-1}$ per H$_2$. The dominant effect on the abundances of both C$_2$H and HCN is the C/O ratio rather than the cosmic ray ionization rate.  \label{fig:2dabun}}
\end{centering}
\end{figure*}

\begin{figure*}
\begin{centering}
\includegraphics[width=1\textwidth]{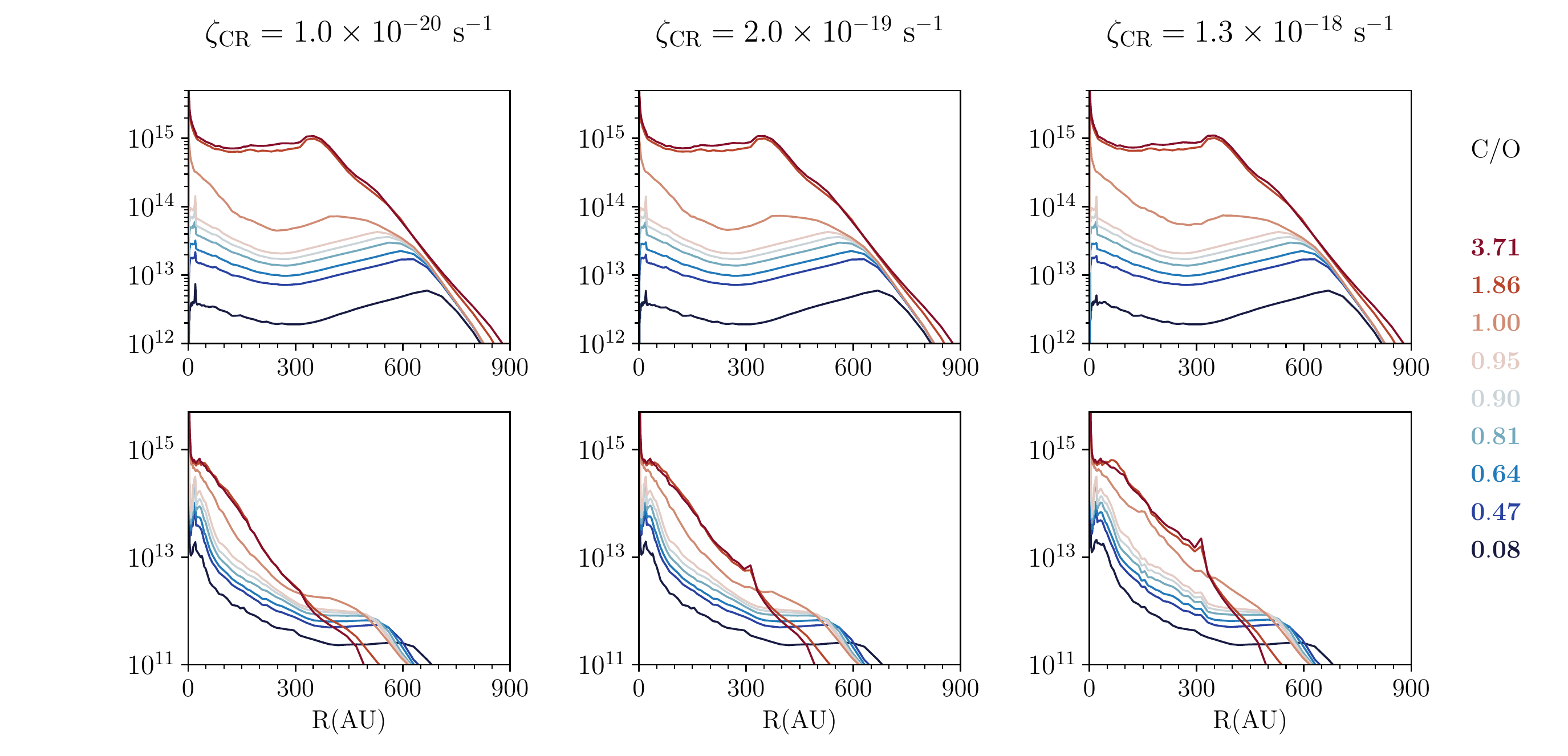}
\caption{C$_2$H (top) and HCN (bottom) chemical model column densities versus radius for different cosmic ray rates as indicated above each column, and at different C/O ratios as indicated by line color on the righthand side. Note the general monotonic trend of decreasing C$_2$H and HCN column density with decreasing C/O.  \label{fig:column}}
\end{centering}
\end{figure*}

Figure~\ref{fig:2dabun} presents the 2D chemical  abundances for a selection of C/O ratios and the intermediate CR rate value. 
It is clear that the C$_2$H is strongly sensitive to the C/O ratio in the gas, confirming early results of \citet{bergin2016}. The C$_2$H abundance is most sensitive for our models below C/O of 1.86, where there is a sharp column density transition straddling C/O of $\sim1$, clearly seen in the two orders of magnitude jump in C$_2$H column densities going from C/O of 1.86 to 0.95 in Figure~\ref{fig:column}.  C$_2$H is essentially unaffected by the cosmic ray ionization rate for the three model values considered. This lack of dependence occurs because C$_2$H is abundant in a layer wrapping around the disk that is UV dominated rather than cosmic ray dominated \citep[see Figure~\ref{fig:2dabun} and][]{bergin2016}.

We can also see that the radial morphology of the column densities changes substantially with C/O (Figure~\ref{fig:column}). For high values $\ge1$, the C$_2$H column density forms a narrow ring or is centrally peaked. At low C/O ratios, the C$_2$H column density forms a wide ring that peaks outside of the millimeter dust disk ($R_{\rm mm}=313$~AU). This transition occurs once the layer of C$_2$H interior to $R\lesssim300$ AU becomes thin, with little column density compared to the outer disk.
Correspondingly, the radial morphology of C$_2$H can change from peaked to ringed even with a uniform C/O ratio, not necessarily requiring (but also not excluding) radial variations in C/O \citep{bergin2016}. 

HCN is sensitive to both the C/O ratio in the gas and the cosmic ray ionization rate, especially for models with C/O $\ge 1$. Higher values of C/O and $\zeta_{\rm CR}$ generally increase the HCN abundance. At a given radius, similar column densities are achieved in models with high CR rates and low C/O, and in models with low CR rates and high C/O, though the overall effect on the radial column density profile is different (Figure~\ref{fig:column}).
The 2D abundance morphologies are similar between HCN and C$_2$H, though we find the HCN abundance is generally less than that of C$_2$H for a given C/O value. One morphological difference is that HCN extends vertically deeper, with a base of $z\sim60$ AU at $R=300$ AU compared to C$_2$H, which disappears below $z\sim80$~AU for C/O $< 1$.

\subsection{Fits to {\rm C$_2$H} Observations}\label{c2hfit}

Based upon the chemical modeling results in Section~\ref{chemresults}, we can use the C$_2$H observations to constrain the C/O ratio in the IM Lup disk's warm molecular layer using the grid of C/O models described in Table~\ref{tab:ini} and illustrated in Figures \ref{fig:2dabun} and \ref{fig:column}. The data and models are compared in the visibility plane with the procedure described in Section~\ref{sec:linemod}.
\begin{figure*}
\begin{centering}
\includegraphics[width=1\textwidth]{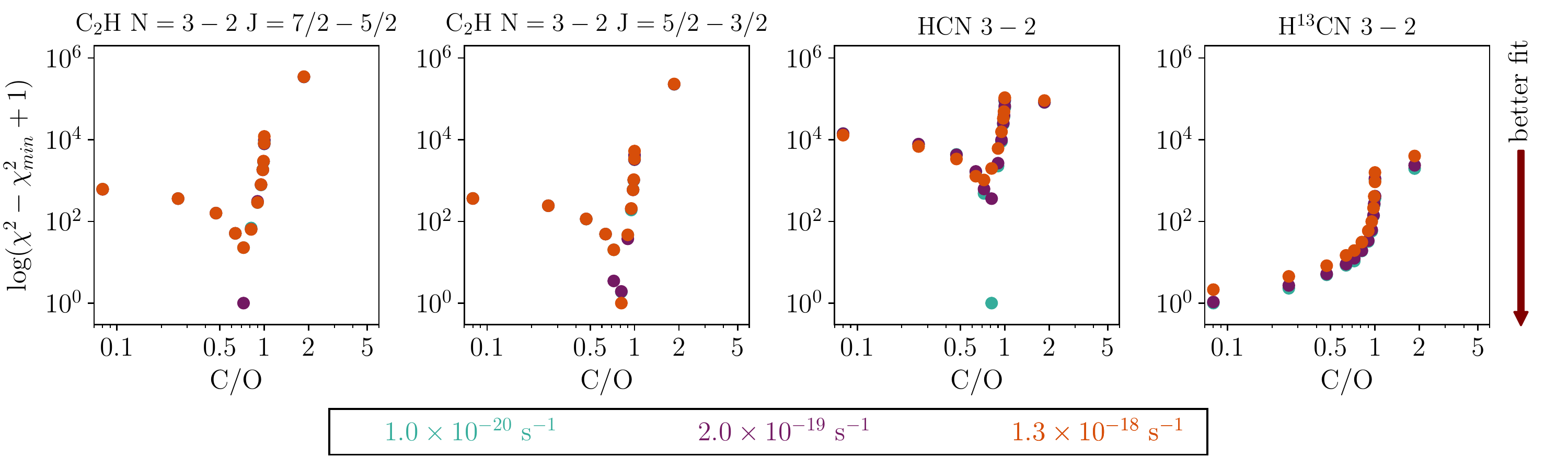}
\caption{$\chi^2$ comparison between the observed visibilities and models for each of the lines or pairs (see top of each panel). Blended hyperfine components for the C$_2$H lines are modeled simultaneously.  From left to right, $\chi^2_{\rm min}$ is $4.58\times10^{6}$, $4.39\times10^{6}$, $6.43\times10^{6}$, and $6.70\times10^{6}$, so in all cases the reduced $\chi^2$ is close to 1, with the minimum being the closet model-data fit. For H$^{13}$CN the models are consistent with non-detection in the image plane for C/O $<1$. Note: the C/O variations reflect the total elemental ratio in the simulations (gas and ice), see Section~\ref{sec:cono} for gas-phase C/O values. These values should be interpreted as surface ($z/r\gtrsim0.2$) constraints, i.e., the region of abundant C$_2$H (Section~\ref{chemresults}). \label{fig:chisq}}
\end{centering}
\end{figure*}

Figure~\ref{fig:chisq} shows the $(\chi^2$ - minimum $\chi^2 + 1)$ between the data visibilities and our model visibilities varying initial gas composition (C/O) and $\zeta_{\rm CR}$, where smaller values indicate better fits. Note, adding 1 allows us to plot the difference on log scale. For the C$_2$H $N=3-2$ $J=\sfrac{5}{2} - \sfrac{3}{2}$ pair of lines a C/O ratio of $\sim0.81$ is favored, while for the C$_2$H $N=3-2$ $J=\sfrac{7}{2} - \sfrac{5}{2}$ pair, C/O of $\sim0.7$ is the best match. Combining both line pairs  favors the $\rm C/O=0.81$ model. We also find that the C$_2$H does not strongly distinguish between cosmic ray ionization rate values, as we expected from the models in Section~\ref{chemresults}. This C/O ratio corresponds to a water ice depletion factor of $50\times$ in the layer C$_2$H is present compared to an interstellar water ice abundance of $8\times10^{-5}$ per H \citep{boogert2015}.

 In the same figure, we also show the HCN $J=3-2$ and H$^{13}$CN $J=3-2$ results for comparison. C/O values of $0.7 - 0.9$ also fit the HCN data reasonably well, while C/O of $\gtrsim1$ predicts {\em detectable} H$^{13}$CN $J=3-2$ image-plane emission, regardless of cosmic ray ionization rate, inconsistent with observations.

\subsection{Fits to {\rm HCN and H$^{13}$CN} Observations}\label{sec:hcnfit}
\begin{figure}[ht!]
\begin{centering}
\includegraphics[width=0.33\textwidth]{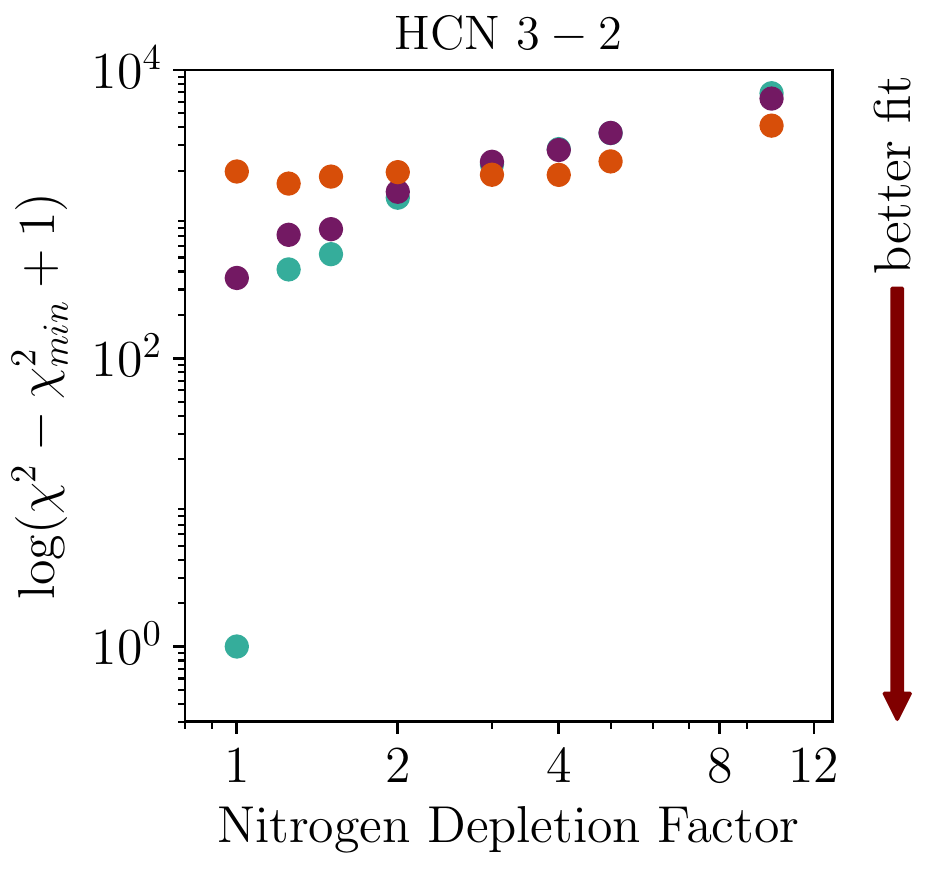}
\caption{$\chi^2$ comparison between the ALMA visibilities and models for HCN $J=3-2$ for differing amounts of initial nitrogen depletion. Colors are the same as in Figure~\ref{fig:chisq} and indicate CR ionization rate. $\chi^2_{\rm min}$ is $6.43\times10^{6}$, so again all models have $\Delta \chi^2$ is close to 1, with smaller values being better fits to the data. For intermediate to low CR rates, no nitrogen depletion is favored. For the higher rate, the models are insensitive to nitrogen depletion factors of $\le 4$ \label{fig:chisqN}}
\end{centering}
\end{figure}

We now consider models with bulk nitrogen depletion to see whether we can arrive at a better fit for HCN $J=3-2$ at a fixed value for C/O. Taking bulk $\rm C/O=0.81$ from Section~\ref{c2hfit}, we reduce the nitrogen abundance from the fiducial value of $7.5\times10^{-5}$ N per H (i.e., $3.75\times10^{-5}$ N$_2$ per H, see Table~\ref{tab:other}). Figure~\ref{fig:chisqN} shows the $(\chi^2$ - $\chi_{\rm min}^2 + 1)$  values for this sub-grid of reduced nitrogen models.  

The global best fit is the low $\zeta_{\rm CR}$ value, $1\times10^{-20}$ s$^{-1}$, and no nitrogen depletion. If instead we compare models within a fixed CR ionization rate, the intermediate CR ionization rate model favors no N-depletion, while the the high $\zeta_{\rm CR}=1.3\times10^{-18}$ s$^{-1}$ model, favors anywhere between no and $4\times$ reduction in bulk nitrogen. However, even this factor is small compared to the $\sim20\times$ depletion of CO, or that implied for water and oxygen not in CO based on the C$_2$H results, i.e., a $\sim50\times$ depletion.

%%%%%%%%%%%%%%%%%%%%%%%%%%%%%%%%%%%%%%%%%%%%%%%%%%%%%%%

\section{Discussion}\label{sec:discussion}

\subsection{Potential mechanisms behind elevated C/O and N/O ratios}

The relative differences in the abundances of the key volatile carriers are consistent with a picture of sequential loss of volatiles from the warm molecular layer. Water is the least volatile, freezing out at relatively high temperatures. Therefore it will be most impacted by the evolution of grains, through growth and settling, removing oxygen from the surface over time \citep[e.g.,][]{hogerheijde2011}. CO plays an active role in gas and grain surface chemistry, and as such it is gradually converted to species like CO$_2$ and CH$_3$OH, all of which can freeze out onto grains and then become depleted from the surface via dust settling \citep{bergin2014far,schwarz2018}. N$_2$, however, is not as chemically active in the disk surface. At cooler temperatures, below the region of CO freeze-out, N$_2$H$^+$ can form directly from N$_2$ and survive. In the surface, in the presence of CO, this formation pathway is hindered, and most of the surface chemistry requires N$_2$ be dissociated before forming other nitrogen-bearing species. As such, nitrogen will be less chemically ``processed'' than the key oxygen and carbon carriers and is expected to stay in its volatile N$_2$ form given its relatively low desorption temperature \citep{oberg2005}. Consequently, sequestration into ice and subsequent settling should theoretically be less effective for nitrogen-bearing species, consistent with our results from the ALMA observations.

\subsection{Gas-phase C/O and N/O constraints}\label{sec:cono}

The model grid presented in Section~\ref{sec:modabun} focuses on the total elemental abundances in both the gas and solid phases, as these are most relevant for the chemical modeling. The relevant values for comparing to observations of exoplanet atmospheres formed via core accretion are the C/O and N/O ratios specifically in the gas phase. Both CO and N$_2$ are primarily in the gas phase where our observations are most sensitive (i.e., the region where C$_2$H emits, $z/r\gtrsim0.2$), while H$_2$O is primarily ice with some gas-phase H$_2$O from UV photo-desorption (also producing OH).

\begin{figure}[]
\begin{centering}
\includegraphics[width=0.43\textwidth]{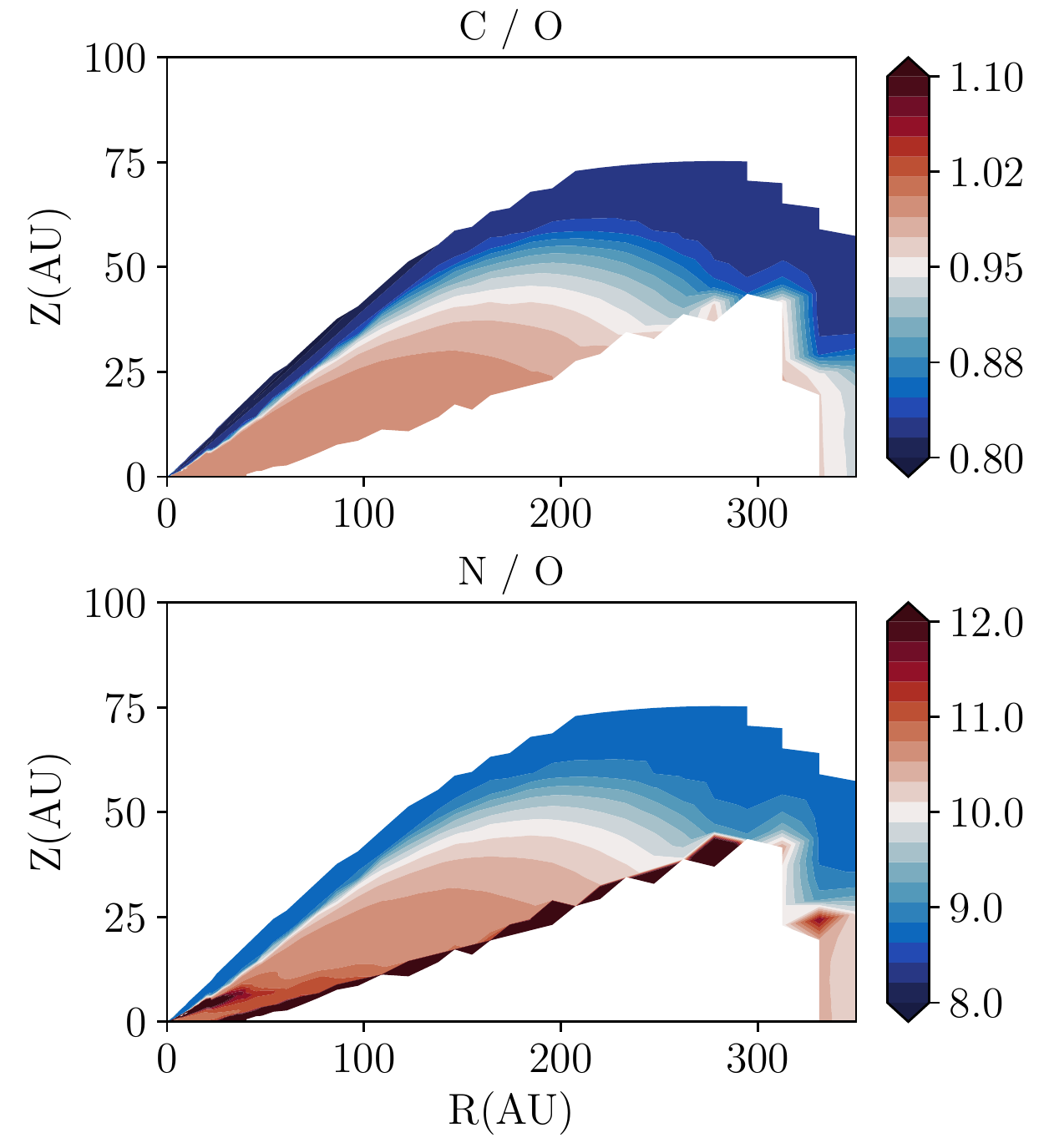}
\caption{ Gas-phase C/O and N/O ratios. Regions that do not have sufficient gas-phase C, O, or N to make a ratio are masked. Note the scale is saturated at the bottom edge of the N/O plot where CO is beginning to freeze out but N$_2$ remains as gas. \label{fig:ratio}}
\end{centering}
\end{figure}

Figure~\ref{fig:ratio} plots the 2D distributions of the {\em gas-phase} C/O and N/O ratios for our best fit model with an intermediate CR ionization rate ($\zeta_{\rm CR}\sim10^{-19}$ s$^{-1}$) and no nitrogen depletion.  Whether we use an intermediate or one order of magnitude lower CR ionization rate does not significantly impact these results.  At the surface, all oxygen that starts in H$_2$O ice is quickly dissociated to gas-phase atomic oxygen, such that the gas C/O ratio is equal to the bulk ratio of $\sim0.8$. Where water is both frozen-out and shielded from UV deeper in the disk, CO becomes  the primary carbon and oxygen carrier in the gas phase, resulting in C/O $\sim1$. The layered C/O structure is visible here, where C/O in the gas is $0.8$ above normalized heights of $z/r>0.3$ and $\sim1$ below this layer until CO starts to freeze-out, at $z/r \lesssim 0.1$. 

The N/O ratio on the other hand is $\gg 1$ throughout the disk atmosphere, and $\sim 10$ when most of the nitrogen is in the gas and CO is the primary oxygen carrier.  In the layer where CO begins to freeze-out, N$_2$ is still in the gas due to its slightly lower binding energy \citep{fayolle2016}, and the gas-phase N/O ratio can be $>100$.

\begin{figure*}
\begin{centering}
\includegraphics[width=1\textwidth]{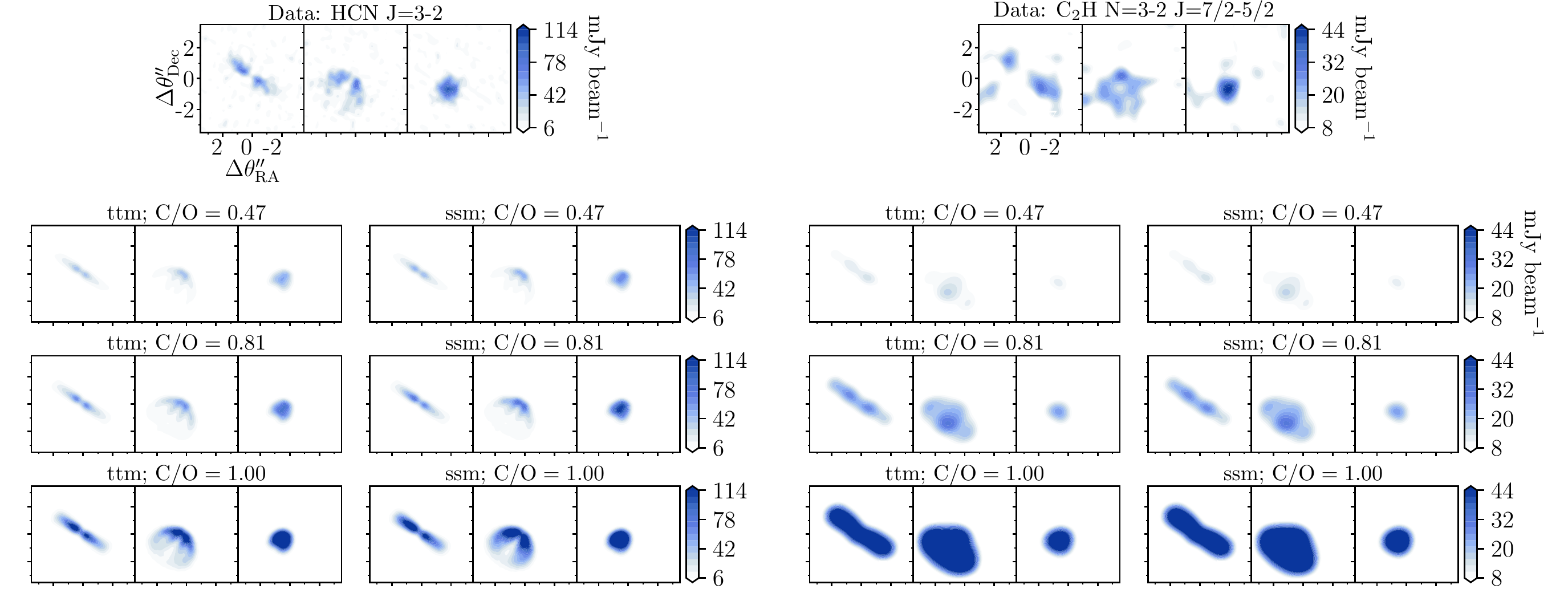}
\caption{Sample model channel maps at the line center and at the line wings for HCN $J=3-2$ (left) and C$_2$H $N=3-2$ $J=\sfrac{7}{2}-\sfrac{5}{2}$ (right) lines, along with the data (top, respectively) for comparison. Rows show varying C/O ratios while columns show the low end and high end of the cosmic ray ionization rate considered.
\label{fig:channelmodel}}
\end{centering}
\end{figure*}

At high velocities we find a discrepancy between the data and modeled C$_2$H, which can be seen in Figure~\ref{fig:channelmodel}. The high velocity emission of C$_2$H is weaker in the models than in the data. Given the signal to noise of the C$_2$H data, we did not attempt  to radially or vertically vary the carbon, oxygen, and nitrogen abundances spatially in our fitting; however, from other sources with varied ring-like morphologies in C$_2$H \citep{bergin2016}, there is support for local variations in bulk C/O, N/O, etc. In this instance, the bright high velocity C$_2$H suggests that the inner disk has more excess carbon than the outer disk, which may point to slower sequestration of carbon-bearing volatiles in this region due to, e.g., reduced freeze-out in the warm inner disk. 

We also examined models where we allowed the disk to extend out to 1000 AU, beyond the $\sim500$~AU radial region where C$_2$H and HCN are observed. Since the model C$_2$H abundance continues to rise in the outer disk for the $\rm C/O=0.81$ case (see Figure~\ref{fig:2dabun}), we predict some emission beyond 500 AU, which was not seen. There may be a change in the gas surface density profile near this radius \citep{cleeves2016im,avenhaus2018}, or the disk may have a lower C/O ratio in this region, i.e., may have less efficient volatile sequestration at these very low disk densities. Higher signal to noise, resolved observations of C$_2$H toward this source will help disentangle these scenarios.

\section{Conclusions}\label{sec:conclusion}

Using detailed physical and chemical models constrained by CO and dust observations, along with C$_2$H and HCN data from ALMA, we constrain the C/O and N/O ratios in the molecular layer of the IM Lup disk. Our observations trace the properties of the disk where C$_2$H and HCN emit, primarily at normalized heights above $z/r\gtrsim0.2$. In these layers, the C$_2$H observations favor a super-solar elemental C/O ratio of $\sim0.8$. This high ratio is consistent with preferential loss of water ice from the surface, e.g., due to sequestration by an evolving and growing population of grains. We do not need to sequester any nitrogen for our best fit model, such that the N/O ratio is also super-solar, $\sim10$.  The gas phase values of C/O and N/O vary spatially depending on the degree of UV shielding at a given location (Section~\ref{sec:cono}). 

While grain sequestration of ices tends to remove volatiles from the surface, it implicitly carries them to the midplane and into the inner disk through settling and radial drift \citep[e.g.][]{oberg2016,piso2016,krijt2016}. If this interpretation is correct, this process would result in a large enhancement over interstellar values of water ice in the inner disk midplane, a mild enhancement of carbon-bearing ice, and relatively little nitrogen ice transport in the solid phase. Correspondingly, we expect the C/O and N/O ratios in midplane solids to both be lower than solar, with N/O much less than solar in the disk midplane. 

Where settled ices eventually end up radially is still an open question. Radial drift of solids is thought to be quite efficient \citep[see review of][]{testi2014}, but this process may be slowed by the emergence of pressure variations in the disk that can effectively trap solids \citep[e.g.,][]{weidenschilling1980}. Now with ALMA, ringed radial structures are being observed, and may even be common \citep[e.g.,][]{alma2015,andrews2016,isella2016,loomis2017,huang2018,fedele2018}. In the absence of pressure traps, these grains should travel inward, thermally desorb, and enhance primarily oxygen, followed by carbon, and relatively little nitrogen. \citet{salyk2011} indeed found very low N/O ratios, $5\times10^{-4}$, in the inner disk with {\em Spitzer}; however, the nitrogen ``correction factors'' in the inner disk gas from HCN to total N are uncertain and require additional chemical modeling to constrain.

These results have interesting consequences for the debate regarding abundance measurements in disks \citep[see summary of][]{bergin2017}. At the low gas temperatures typical of disks, H$_2$ does not emit appreciably \citep{bergin2013hd,mcclure2016,bergin2017}. As a result, other uncertain mass tracers are typically used, such as the total dust mass multiplied by a conversion factor, or even optically thin CO emission itself. From our modeling, we find that we do not need to deplete nitrogen significantly, with a factor of $4-20\times$ difference between the CO depletion factor and that for nitrogen. These results would point to abundance variations between these species rather than an overall under-accounting of disk mass, which would impact all volatile abundances at similar if not equal levels. We of course cannot rule out with these data alone some missing gas mass, since for the higher CR models, a small amount of nitrogen depletion (a factor of a few) is allowed, even though these are not the global best fits. However, missing gas alone cannot explain the higher degree of depletion needed for CO and water ice. In the future, additional observations of nitrogen-bearing molecules like HCN may help to break this CO mass / gas mass degeneracy, where in larger disk surveys CO masses appear globally low relative to the dust masses scaled by the interstellar gas-to-dust ratio \citep{ansdell2016,miotello2016,long2017}.

Going forward, these results show that readily observable molecular tracers like CO, C$_2$H, and HCN, and their isotopologues combined with astrochemical models can be used to constrain the gas-phase C/O and N/O ratios within planet-forming disks. Such measurements may furthermore shed light on the inferred volatile composition of the ices, once we better understand the mechanism(s) of volatile loss from the warm molecular layer \citep[e.g.,][]{furuya2014,kama2016}. Future  high sensitivity, resolved observations  of many sources may further help shed light on ``typical'' gas-phase C/O and N/O ratios, how they spatially vary, and how these ratios may vary with time as planets are forming in the disk. Together, these can one day be compared with C/O measurements of exoplanet atmospheres, to eventually help unravel their formation locations and histories, where observations have already shown a wide range of exoplanet C/O values, even within a single planetary system \citep{bonnefoy2016,lavie2017}.

\acknowledgements{{\it Acknowledgements:} This paper makes use of the following ALMA data: ADS/JAO.ALMA\#2015.1.00964.S and ADS/JAO.ALMA\#2013.1.00226.S. ALMA is a partnership of ESO (representing its member states), NSF (USA) and NINS (Japan), together with NRC (Canada) and NSC and ASIAA (Taiwan) and KASI (Republic of Korea), in cooperation with the Republic of Chile. The Joint ALMA Observatory is operated by ESO, AUI/NRAO and NAOJ. The National Radio Astronomy Observatory is a facility of the National Science Foundation operated under cooperative agreement by Associated Universities, Inc. LIC acknowledges the support of NASA through Hubble Fellowship grant HST-HF2-51356.001-A awarded by the Space Telescope Science Institute, which is operated by the Association of Universities for Research in Astronomy, Inc., for NASA, under contract NAS 5-26555. J.H. acknowledges support from the National Science Foundation Graduate Research Fellowship under Grant No. DGE-1144152. All simulations were carried out using the Smithsonian Institution High Performance Cluster (SI/HPC). 
}

%\bibliography{ms}

\providecommand{\od}{O'D} \providecommand{\accO}{\'O}

\appendix
\section{Updated grain surface chemistry}\label{app:grains}

The amount of dust surface area impacts the rate of grain surface chemistry that can occur. Dust surface area can be altered by removing dust grains (but keeping their size spectrum constant), or changing the size spectrum. The default used in the chemical code is 0.1 micron-sized grains with an abundance of $\chi_{\rm gr}^0=6\times10^{-12}$ per H atom \citep{fogel2011}. The reason for this default size, which is bigger than the minimum grain size used to compute the temperature structure, is because grains smaller than this can be heated by single-photon events important below $0.06\mu$m \citep{leger1985}, and thus have stochastic chemical behavior not adequately described by the present treatment in the code.

In the sweeping approximation for surface chemistry \citep[e.g.][]{hhl}, the rates of surface reactions are given by:

\begin{equation}
R_{i,j}  = b_{i,j} \left(R_{\rm diff,i} + R_{\rm diff,j} \right) N_i N_j n_{\rm gr},
\end{equation}
where $N_x$ represents the total number of species $x$ on the average grain and $n_{\rm gr}$ is the number density of dust grains. The dimensions of $R_{i,j}$ are per volume per time. The $b_{i,j}$ term factors in for reactions with a barrier, where the probability for tunneling can be approximated by: 
\begin{equation}
b_{i,j}  = \exp\left[ -2 (a / \hslash) \sqrt{2 \mu E_a} \right],
\end{equation}
where $a$ is the area of a site taken to be 1 \AA, and $E_a$ is the activation energy of a given reaction. The rate coefficient $\kappa_{i,j}$ for the grain-surface reaction of species $i$ and $j$ is then:
\begin{align}
\kappa_{i,j} & = \frac{R_{i,j} }{n_i n_j} \\
 & =  \frac{b_{i,j} \left(R_{\rm diff,i} + R_{\rm diff,j} \right) N_i N_j n_{\rm gr}}{n_i n_j} \\
& = \frac{b_{i,j} \left(R_{\rm diff,i} + R_{\rm diff,j} \right) \frac{n_i}{n_{\rm gr}} \frac{n_j}{n_{\rm gr}} n_{\rm gr}}{n_i n_j} \\
& = {b_{i,j} \left(R_{\rm diff,i} + R_{\rm diff,j} \right) \frac{1}{n_{\rm gr}}  }
\end{align}

For grains where there is a substantial ice mantle, only the surface monolayer(s) should participate in the chemistry, such that the $N_i$ and $N_j$ terms should be reduced by the ratio of the volume density of ice in the reactive surface area, $N_{\rm sites} n_{\rm ice}$, to the total volume density of ice, $n_{\rm ice}$, such that the dilution $d$ is given by:
\begin{equation}
d = \frac{n_{\rm gr} N_{\rm sites} }{n_{\rm ice}},
\end{equation}
where $N_{\rm sites}$ is the number of surface sites per grain. 
The numerator, $n_{\rm gr} N_{\rm sites}$, is the number of sites per volume, which equals $n_{\rm gr}  \sigma_{\rm gr} / \sigma_{\rm site}$.

We define $\Pi_{\rm adj}$ to represent the ratio of adjusted surface area per volume due to grain removal or growth compared to the standard case of 0.1 $\mu$m sized grains:
\begin{align}
\Pi_{\rm adj} & = \frac{\sigma_{\rm gr} n_{\rm gr}}{ \sigma_{\rm gr}^0 n_{\rm gr}^0}.
\end{align}
If we assume an MRN grain size distribution \citep{mrn1977} where the number of grains $n_{\rm gr}$ is proportional to the size of the grains to $r^{-3.5}$, we can calculate the population integrated quantity $n_{\rm gr}  \sigma_{\rm gr}$ to be:
\begin{equation}
n_{\rm gr}  \sigma_{\rm gr} = -3  \rho_d   / \rho_{si}  \left[\frac{r_{\rm max}^{-0.5}-r_{\rm min}^{-0.5}} {r_{\rm max}^{0.5} - r_{\rm min}^{0.5}}\right],
\end{equation}
where $r_{\rm min}$ is 0.06~$\mu$m, i.e., the single-photon heating limit of \citet{leger1985}  and $r_{\rm max}$ varies. $\rho_d$ is the volumetric density of dust and $\rho_{si}$ is the density of silicates. Specifically, we assume two distinct dust populations, ``big'' grains that have an upper size limit of $r_{\rm max}=1$~mm, and ``small'' grains that have an upper size limit of  $r_{\rm max}=1$~$\mu$m as described in \citet{cleeves2016im}. Taking into account both populations, and substituting this expression into $\Pi_{\rm adj}$ gives:
\begin{align}
\Pi_{\rm adj} & =  
100 \frac{ \rho_d }{\rho_{\rm gas}} \frac{  f_{\rm big}  \left[\frac{r_{\rm max, big}^{-0.5} - r_{\rm min}^{-0.5}} {r_{\rm max, big}^{0.5} - r_{\rm min}^{0.5}}\right] +  f_{\rm sm}  \left[\frac{r_{\rm max, sm}^{-0.5} - r_{\rm min}^{-0.5}}{r_{\rm max, sm}^{0.5} - r_{\rm min}^{0.5})}\right]  }{  \left[ \frac{r_{\rm max, sm}^{-0.5} - r_{\rm min}^{-0.5}}{r_{\rm max, sm}^{0.5} - r_{\rm min}^{0.5}}\right]}.
\end{align}
From this, we can now express the grain surface area per unit volume by 
\begin{equation}
\sigma_{\rm gr} n_{\rm gr} = \Pi_{\rm adj} \sigma_{\rm gr}^0 n_{\rm gr}^0  = \Pi_{\rm adj} \sigma_{\rm gr}^0  \chi_{\rm gr}^0 n_{\rm H} 
\end{equation}
allowing us to express the dilution term, $d$, as:
\begin{equation}
d = \frac{ \Pi_{\rm adj} \sigma_{\rm gr}^0  \chi_{\rm gr}^0 n_{\rm H}  / \sigma_{\rm site}}{n_{\rm ice}}.
\end{equation}
In the case where both $i$ and $j$ are spread over the full ice mantle (i.e., are diluted), the rate decreases by $d^2$:
\begin{align}
\kappa_{i,j} &=  d^2  \frac{b_{i,j} \left(R_{\rm diff,i} + R_{\rm diff,j} \right)}{n_{\rm gr}}  \\
&  =  \left({\frac{ \Pi_{\rm adj} \sigma_{\rm gr}^0  \chi_{\rm gr}^0 n_{\rm H}}{\sigma_{\rm site} n_{\rm ice}}}\right)^2  \frac{b_{i,j} \left(R_{\rm diff,i} + R_{\rm diff,j} \right)}{n_{\rm gr}}.
\end{align}
The final piece we need is to estimate $n_{\rm gr}$ in terms of $\Pi_{\rm adj}$. We can do this approximately by rearranging our expression for $\sigma_{\rm gr} n_{\rm gr}$:
\begin{align}
 n_{\rm gr} = \frac{\Pi_{\rm adj} \sigma_{\rm gr}^0 n_{\rm gr}^0 }{ < \sigma_{\rm gr}>}, 
 \end{align}
 and then calculate the average grain cross section again by computing the weighted average $\sigma$ with the MRN grain size distribution:
\begin{align}
<\sigma_{\rm gr} > &  = 5 \pi \left[ \frac{r_{\rm max}^{-0.5}-r_{\rm min}^{-0.5}}{r_{\rm max}^{-2.5}-r_{\rm min}^{-2.5}}\right] \\
    & \approx 5 \pi \left[ \frac{-r_{\rm min}^{-0.5}}{-r_{\rm min}^{-2.5}}\right] = 5 \pi r_{\rm min}^{2}.
 \end{align} 
Using this expression, we estimate the grain number density to be:
  \begin{align}
  n_{\rm gr} 
  & =  \frac{ \Pi_{\rm adj} \sigma_{\rm gr}^0 \chi_{\rm gr}^0 n_{\rm H} }{ 5 \pi r_{\rm min}^{2} } \\
   & \approx \frac{1}{1.8} \Pi_{\rm adj} \chi_{\rm gr}^0 n_{\rm H} 
\end{align}
where the prefactor comes from $\sigma_{\rm gr}^0  = \pi (0.1\mu m)^2$, $r_{\rm min} = 0.06 \mu m$. 
We can now solve for the rate coefficient $\kappa$: 
\begin{align}
\kappa_{i,j} & =  \left({\frac{ \Pi_{\rm adj} \sigma_{\rm gr}^0  \chi_{\rm gr}^0 n_{\rm H}}{\sigma_{\rm site} n_{\rm ice}}}\right)^2  {b_{i,j} \left(R_{\rm diff,i} + R_{\rm diff,j} \right)}\frac{ 5 \pi r_{\rm min}^{2} }{ \Pi_{\rm adj} \sigma_{\rm gr}^0 \chi_{\rm gr}^0 n_{\rm H} } \\
& = \frac{N_{\rm sites}^2}{n_{\rm ice}^2} \Pi_{\rm adj} \chi_{\rm gr}^0 n_{\rm H}  {b_{i,j} \left(R_{\rm diff,i} + R_{\rm diff,j} \right)} \frac{5 \pi r_{\rm min}^{2}}{ \sigma_{\rm gr}^0}.
\end{align}
In this case, for decreasing surface area per unit volume $\Pi_{\rm adj}$, the reaction rate decreases since the amount of reactive ice (i.e., that on the surface and not buried in the mantle) goes down when there is less surface area for freeze out and individual mantles become thicker.

For the case of one diluted species and one not (like reactions with atomic hydrogen, where we assume that any atomic hydrogen is part of the reactive surface) we only multiply the rate by one dilution $d$ factor,
\begin{align}
\kappa_{i,j} & =  \left({\frac{ \Pi_{\rm adj} \sigma_{\rm gr}^0  \chi_{\rm gr}^0 n_{\rm H}}{\sigma_{\rm site}  n_{\rm ice}}}\right)  b_{i,j} \left(R_{\rm diff,i} + R_{\rm diff,j} \right)  \frac{ 5 \pi r_{\rm min}^{2} }{ \Pi_{\rm adj} \sigma_{\rm gr}^0 \chi_{\rm gr}^0 n_{\rm H} } \\
& = {\frac{N_{\rm sites}}{n_{\rm ice}}}  b_{i,j} \left(R_{\rm diff,i} + R_{\rm diff,j} \right)  \frac{5 \pi r_{\rm min}^{2}}{\sigma_{\rm gr}^0}.
\end{align}
The dependence on $\Pi_{\rm adj}$  drops out because the diluted species is penalized by more inert ice locked up in the non-reactive mantle, while the other more mobile species gains from $\Pi_{\rm adj}$ because there are fewer possible grains to occupy, so any single grain harbors more of the undiluted hydrogen atoms.
For the case where neither species is diluted, like H(gr) + H(gr) $\rightarrow$ H$_2$,
\begin{align}
\kappa_{i,j} & =  b_{i,j} \left(R_{\rm diff,i} + R_{\rm diff,j} \right)  \frac{ 5 \pi r_{\rm min}^{2} }{ \Pi_{\rm adj} \sigma_{\rm gr}^0 \chi_{\rm gr}^0 n_{\rm H} },
\end{align}
the rate now increases with decreasing grain surface area per unit volume $\Pi_{\rm adj}$ for the same reason as the case above, i.e., the probability of two H-atoms existing on the same grain and reacting to form H$_2$ goes up when  there are fewer possible grains for them to occupy.
The main result of incorporating these reactions is an overall decrease in the efficiency grain-surface chemistry for molecules other than molecular hydrogen, where before under certain conditions CO would be quickly converted into CO$_2$, H$_2$CO, or CH$_3$OH ice. These species still form, but at a reduced abundance, a few percent up to tens of percent of the CO ice abundance just at or below the CO snow surface. 

\end{document}